\documentclass[aps,prb,onecolumn,amsmath,amssymb,superscriptaddress]{revtex4}
\usepackage{graphicx}
\usepackage{amssymb}
\usepackage{natbib}
\usepackage[compact]{titlesec}
\usepackage[rawfloats=true]{floatrow}
\usepackage{array}
\usepackage{chngcntr}
\restylefloat{figure}

\begin{document}

\title{Magnetic order and fluctuations in quasi-two-dimensional planar magnet Sr(Co$_{1-x}$Ni$_{x})_{2}$As$_{2}$}
\author{Yaofeng Xie}
\affiliation{Department of Physics and Astronomy, Rice University, Houston, Texas 77005, USA}

\author{Yu Li}
\email{yuli1@lsu.edu}
\affiliation{Department of Physics and Astronomy, Louisiana State University, Baton Rouge, Louisiana 70803, USA}

\author{Zhiping Yin}
\email{yinzhiping@bnu.edu.cn}
\affiliation{Center for Advanced Quantum Studies and Department of Physics, Beijing Normal University, Beijing, China}

\author{Rui Zhang}
\author{Weiyi Wang}
\affiliation{Department of Physics and Astronomy, Rice University, Houston, Texas 77005, USA}

\author{Matthew B. Stone}
\author{Huibo Cao}
\author{D. L. Abernathy} 
\affiliation{Neutron Scattering Division, Oak Ridge National Laboratory, Oak Ridge, Tennessee 37831, USA}

\author{Leland Harriger}
\affiliation{NIST Center for Neutron Research, National Institute of Standards and Technology, Gaithersburg, Maryland 20899, USA}

\author{David P. Young}
\author{J. F. DiTusa}
\affiliation{Department of Physics and Astronomy, Louisiana State University, Baton Rouge, Louisiana 70803, USA}

\author{Pengcheng Dai}
\email{pdai@rice.edu}
\affiliation{Department of Physics and Astronomy, Rice University, Houston, Texas 77005, USA}

\begin{abstract}
We use neutron scattering to investigate spin excitations in Sr(Co$_{1-x}$Ni$_{x})_2$As$_2$, which has a $c$-axis incommensurate
helical structure of the two-dimensional (2D) in-plane ferromagnetic (FM) ordered layers for $0.013\leq x \leq 0.25$.
By comparing the wave vector and energy dependent spin excitations 
in helical ordered Sr(Co$_{0.9}$Ni$_{0.1}$)$_2$As$_2$ and paramagnetic SrCo$_2$As$_2$, we find that Ni-doping, while increasing lattice disorder in Sr(Co$_{1-x}$Ni$_{x})_2$As$_2$, 
enhances quasi-2D FM spin fluctuations.
However, our band structure calculations within the combined density functional theory
and dynamic mean field theory (DFT+DMFT) failed to generate a correct incommensurate wave vector for the observed helical order from nested Fermi surfaces. Since transport measurements reveal increased in-plane and $c$-axis electrical 
resistivity with increasing 
Ni-doping and associated lattice disorder, we conclude that 
the helical magnetic 
order in Sr(Co$_{1-x}$Ni$_{x})_2$As$_2$ may arise from  
a quantum order-by-disorder mechanism through the itinerant electron mediated Ruderman-Kittel-Kasuya-Yosida (RKKY) interactions. 
\end{abstract}
%\pacs{}

\maketitle

% Introduction
\section{Introduction}
Two-dimensional (2D) magnetism has been studied by condensed matter physicists for decades. In 1966, Mermin and Wagner showed rigorously that 
thermal fluctuations destroy the 2D long-range magnetic order 
at finite temperature in spin-rotational invariant systems with short-range magnetic interactions describable by a 3D Heisenberg Hamiltonian \cite{WM}.
On the other hand, the exact solution of the 2D spin Ising Hamiltonian reveals a finite temperature magnetic
phase transition in a 2D magnet, where the anisotropic spin component opens a gap in the spin-wave spectrum that suppresses the effect of thermal fluctuations \cite{Ising,Onsager}. For planar 2D magnets where spins are confined within the layer 
described by the $XY$ model, although the susceptibility diverges below a finite temperature $T_{KT}$ 
through a Kosterlitz and Thouless transition \cite{KT}, spin correlations are characterized by 
an algebraic decay with quasi-long-range magnetic order instead of a true long-range order \cite{Rev2D1,Rev2D2}. 
 
In quasi-2D bulk van der Waals 
magnetic materials such as CrI$_3$ \cite{CrI3_FM,Ferro2D_2,Chen2020}, 
Cr$_2$Ge$_2$Te$_6$ \cite{Cr2Ge2T6_FM}, and MnBi$_2$Te$_4$ \cite{MnBi2Te4}, 
a small spin-orbit-coupling induced magnetic anisotropy or
interactions along the $c$-axis can result in a 3D magnetic order.
Since the $c$-axis magnetic exchange interactions depend on the layer thickness and/or stacking, and can be either ferromagnetic (FM) or antiferromagnetic (AF) \cite{Cr2Ge2T6_FM,Chen2020,CrI3_FM,MnBi2Te4,Ferro2D_2},
a detailed investigation of the relationship between the 3D magnetic order and $c$-axis exchange couplings in quasi-2D materials will provide important information on the spin Hamiltonian that governs the magnetic properties of the system.

\begin{figure}[t]
\includegraphics[scale=.5]{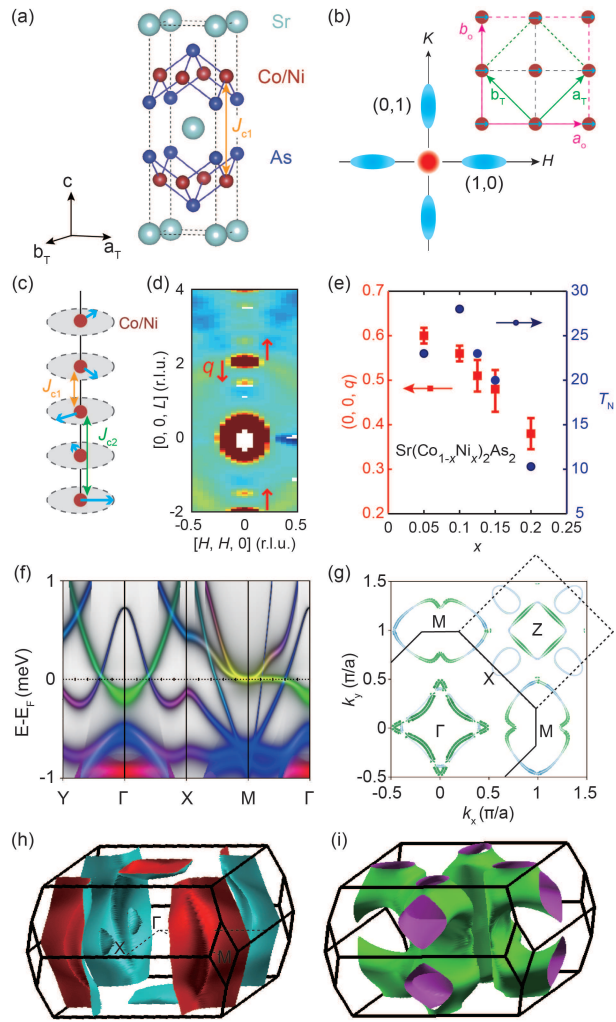}
\caption{
(a) Crystal structure of Sr$($Co$_{1-x}$Ni$_{x}$)$_{2}$As$_{2}$.
(b) Schematics of spin fluctuations in reciprocal space and the 
Co lattice in real space. Red (blue) areas represent FM (AF) spin fluctuations.
(c) Helical magnetic structure in the $x=0.1$ compound.
(d) Neutron diffraction intensity in the $[H,H,L]$ plane. 
The red arrows indicate the helical wave vector.
(e) Magnetic wave vectors and 3D transition 
temperatures in Sr(Co$_{1-x}$Ni$_{x})_2$As$_2$.
(f,g) Electronic band structures and the Fermi surfaces 
for $k_z = 0$ in $x=0.1$. Green (red) color represents the $d_{x^2-y^2}$ ($d_{z^2}$) orbital and blue is the contribution from the $t_{2g}$ ($d_{xz}$, $d_{yz}$, $d_{xy}$) orbitals. Yellow is the mixture of the red ($d_{z^2}$) and green ($d_{x^2-y^2}$).
(h,i) Calculated 3D Fermi surfaces of the $x=0.1$ compound.
}
\end{figure}

Recently, the Co-based $A$Co$_2$X$_2$ ($A =$ Ca, Sr, Ba; $X =$ P,As) system [Fig.1(a)] has attracted considerable attention due to its complex magnetic behavior and close relationship with iron pnictide 
superconductors \cite{CCA2,CSCA,SCA1,Sangeetha17,Dai2015}. 
CaCo$_{1.86}$As$_2$ has strong Stoner-enhanced 2D ferromagentism with the $A$-type AF stacking of the FM CoAs layers with the 
ordered moment aligned along the $c$-axis \cite{CCA2}. When Ca in CaCo$_{2-y}$As$_2$ is partially substituted by Sr to
form Ca$_{1-x}$Sr$_x$Co$_{2-y}$As$_2$, the single-ion spin anisotropy tunes the $c$-axis easy-axis in 
CaCo$_{1.86}$As$_2$ to easy-plane. In addition, the AF propagation vector suddenly changes from $(0,0,1)$ in CaCo$_{1.86}$As$_2$ into $(0,0,0.5)$ in the intermediate doped compounds which corresponds to a periodicity of four CoAs layers \cite{CCA2,CSCA}.
Since both FM and stripe-type AF spin fluctuations are present within the CoAs plane [Fig. 1(b)] of $A$Co$_2$X$_2$ \cite{SCA1,CCA1,SCA_ly}, the subtle balance and competition of the associated FM and AF interactions are responsible 
 for the $A$-type AF order in CaCo$_{2-y}$As$_2$ \cite{CCA2} and 
paramagnetic state without magnetic order 
down to 0.05 K in SrCo$_2$As$_2$ \cite{SCA1}.
While SrCo$_2$P$_2$ and SrCo$_2$Ge$_2$ are both paramagnetic, 
a FM quantum critical point and weak itinerant FM order were induced in SrCo$_2$(Ge$_{1-x}$P$_x$)$_2$ for $0.3<x<0.7$ \cite{SCGP}.  Furthermore, doping electrons into SrCo$_2$As$_2$ by replacing Sr with La/Nd
drives the system into a FM ordered state, thus suggesting 
that SrCo$_2$As$_2$ is close to a FM instability \cite{SLCA,SNCA}. In contrast, 
electron-doping SrCo$_2$As$_2$ by substituting Co with Ni to form 
Sr(Co$_{1-x}$Ni$_x$)$_2$As$_2$ results 
in a helical ordered state with spins aligned ferromagnetically within the layers, but rotating along the $c$-axis 
for $0.05\leq x\leq0.35$ [Figs. 1(c,d)] \cite{SCNA_ly,SCNA_ames,Wilde2019}.

\begin{figure}[t]
    \centering
    \includegraphics[scale=0.6]{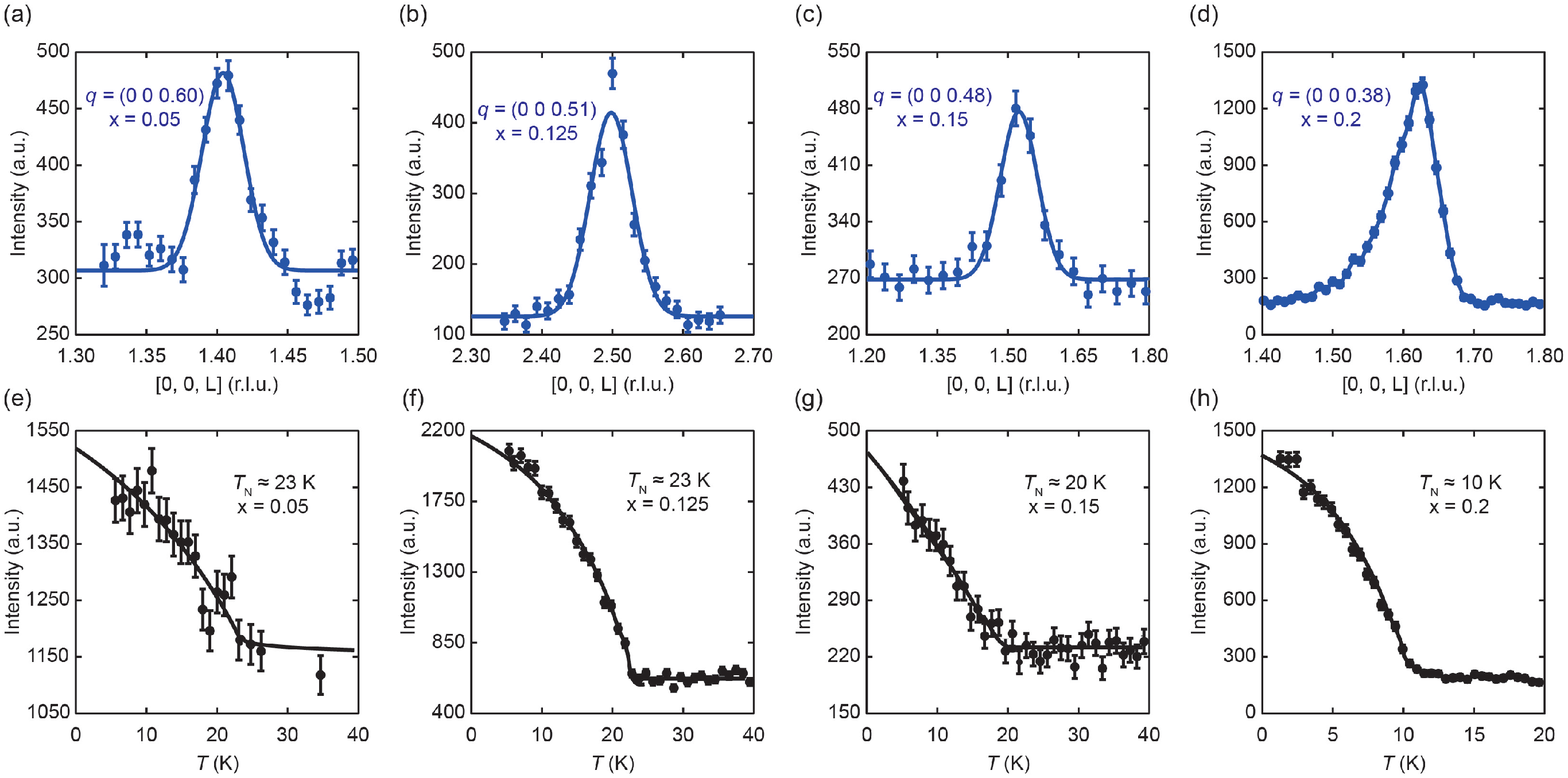}
    \caption{Neutron diffraction results for Sr$($Co$_{1-x}$Ni$_{x})_2$As$_2$ single crystals. (a)-(d) Scan along the $[0,0,L]$ direction for $x = 0.05, 0.125, 0.15,$ and 0.2, respectively. (e)-(h) Temperature dependence of the magnetic order parameter, where solid lines are guides to the eye.}
    \label{fig:my_label}
\end{figure}

Although the helical order in Sr(Co$_{1-x}$Ni$_x$)$_2$As$_2$ can be phenomenologically fit by an 1D frustrated Heisenberg Hamiltonian with an easy-axis spin anisotropy, it should have large  
itinerant electron contributions as reflected by the large Rhodes-Wholfarth ratio \cite{SCNA_ly,SCNA_ames}. 
The presence of a saddle-like flat band nearby the Fermi level results in a high density of electronic states, which should affect the low energy magnetic properties \cite{BCA,ACS_zhiping}. 
Since the flat portion of the band structure near the Fermi level can provide  
 enormous soft particle-hole excitations which can couple to the magnetic order parameter 
and modify the original phase diagram \cite{QOBD,fb1}, a determination of the evolution of 
the magnetic order and spin excitations in Sr(Co$_{1-x}$Ni$_x$)$_2$As$_2$ will form the basis to unveil the microscopic origin of the magnetism in these materials. 

\begin{figure}[t]
\includegraphics[scale=.46]{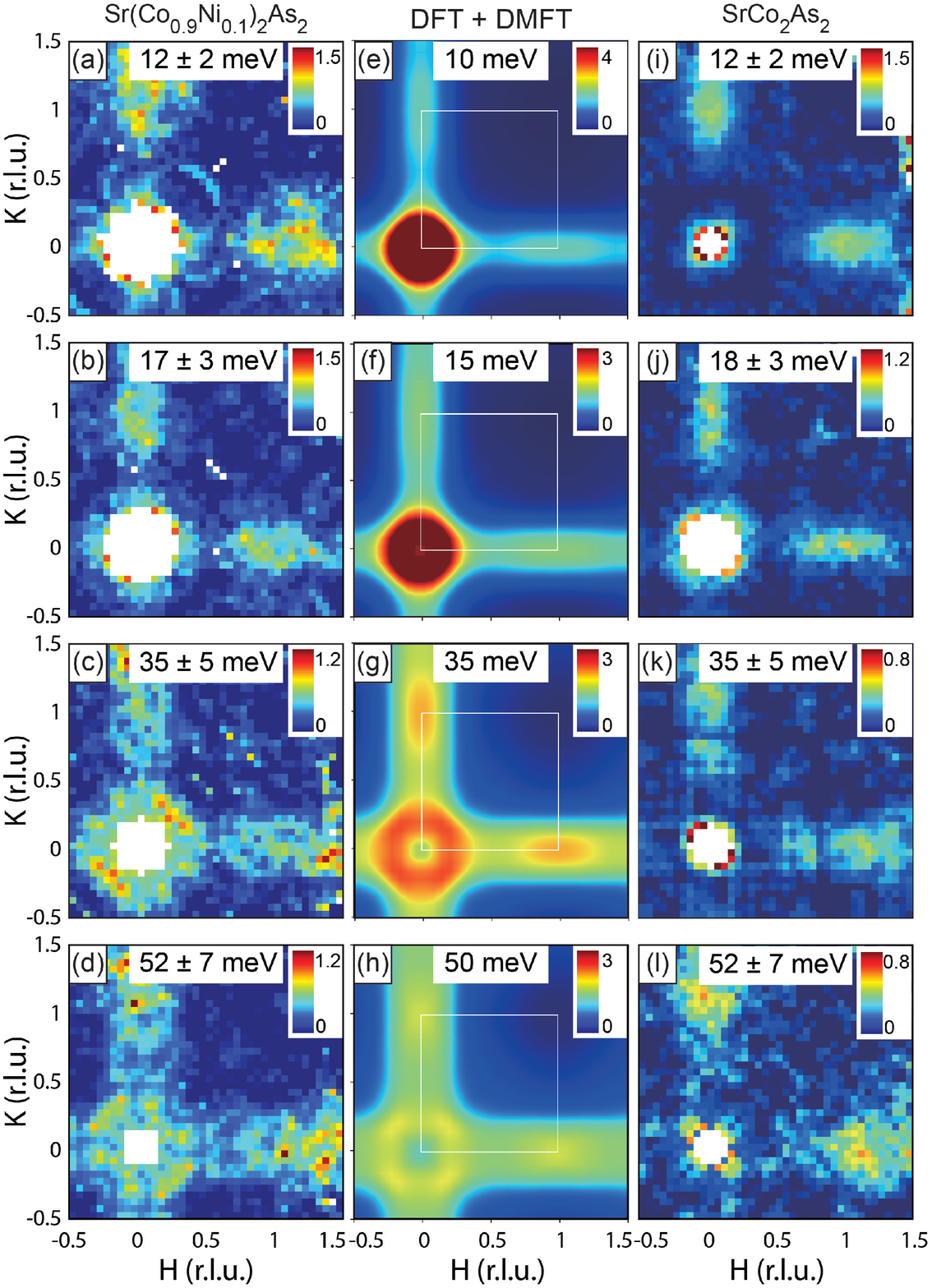}
\caption{(a,b,c,d) Constant energy slices of spin fluctuations of Sr$($Co$_{0.9}$Ni$_{0.1})$As$_2$ within the $[H,K]$ plane at $E = 12 \pm 2, 17 \pm 3, 35 \pm 5$, and $52 \pm 7$ meV, respectively. (e,f,g,h) The dynamic spin susceptibility from the DFT+DMFT calculations at the corresponding energies for Sr$($Co$_{0.9}$Ni$_{0.1})_2$As$_2$. (i,j,k,l) The same constant energy slices in pure SrCo$_2$As$_2$.
}
\end{figure}

\section{Results}

Here we report neutron diffraction studies on the evolution of magnetic structures in a series of Sr(Co$_{1-x}$Ni$_x$)$_2$As$_2$ compounds and inelastic neutron scattering measurements on magnetic excitations in Sr(Co$_{0.9}$Ni$_{0.1}$)$_2$As$_2$ and SrCo$_2$As$_2$. While previous work has established the presence of incommensurate helical magnetic order and  
the Ni-doping dependent phase diagram of Sr(Co$_{1-x}$Ni$_x$)$_2$As$_2$ \cite{SCA_ly,SCNA_ly,SCNA_ames,Wilde2019}, there are no detailed determination of the Ni-doping dependent magnetic structure and spin excitations.  
 We find that the magnetic propagation vector evolves from $q = 0.6$ in $x=0.05$ to $q$ = 0.38 in $x=0.2$, suggesting that the $T_N = 0$ in Sr(Co$_{1-x}$Ni$_x$)$_2$As$_2$ with $x\approx 0.25$ may be driven by FM fluctuations.  Based on our inelastic neutron scattering results on Sr(Co$_{0.9}$Ni$_{0.1}$)$_2$As$_2$ and SrCo$_2$As$_2$, we conclude that the Ni substitution into SrCo$_2$As$_2$ induces strong FM magnetic excitations which are 2D and dispersionless along the $c$-axis direction in the measured energy range. 
By comparing the magnetic ordering wave vectors with the band structures calculated using
the combined density functional theory
and dynamic mean field theory (DFT+DMFT) \cite{SCA_ly}, we find no conclusive evidence for Fermi surface nesting induced helical order. Since transport measurements reveal increased in-plane and $c$-axis electrical 
resistivity with increasing 
Ni-doping and associated lattice disorder, we conclude that 
the helical magnetic 
order in Sr$($Co$_{1-x}$Ni$_{x})$As$_2$ may arise from  
a quantum order-by-disorder mechanism through the itinerant electron 
mediated Ruderman-Kittel-Kasuya-Yosida (RKKY) interactions \cite{QOBD}.

Single crystals of Sr$($Co$_{1-x}$Ni$_{x})_2$As$_2$ with different Ni-doping levels, $x$, were synthesized from solution using self-flux method with the molar ratio of Sr:NiAs:CoAs = 1:5$x$:5(1-$x$). The starting materials were placed in an alumina crucible and sealed in an evacuated quartz tube. The sealed mixture was first heated slowly to 830 ${^\circ}$C, and then cooked at 1195 ${^\circ}$C for 20 hours. Then the furnace was slowly cooled down to 980 ${^\circ}$C at the rate of 3 ${^\circ}$C/h. Single crystals with typical sizes of 0.5 centimeters were obtained by cleaning off the flux.  To confirm the nominal compositions of the samples, we used inductively coupled plasma mass spectrometry analysis to determine the actual chemical composition of the resulting single crystals.  For this purpose, we looked 8 single crystals with different Ni-doping $x$.  The results are summarized in Table I.  Assuming As concentration to be correct at 100\%, we find that the normal and actual chemical compositions of the samples are very similar, indicating that the nominal Ni-doping level is a good representation of the actual doping level.

\vspace{1cm}
\begingroup
\setlength{\tabcolsep}{10pt} % Default value: 6pt
\renewcommand{\arraystretch}{1.5} % Default value: 1
\begin{table}
\begin{center}
	\begin{tabular}{l l l l l l l}
	\hline
Samples & As & Co & Ni & Sr & Nominal Ni-doping & Actual Ni-doping\\
		\hline
		1  & 1 & 0.78 & 0.19 & 0.5 & 0.2 & 0.2\\
		2  & 1 & 0.78 & 0.19 & 0.5 & 0.2 & 0.2\\
		3  & 1 & 0.81 & 0.14 & 0.49 & 0.15 & 0.15\\
		4  & 1 & 0.84 & 0.13 & 0.5 & 0.15 & 0.13\\
    5  & 1 & 0.89 & 0.09 & 0.47 & 0.1 & 0.09\\
		6  & 1 & 0.86 & 0.09 & 0.5  & 0.1 & 0.09\\
		7  & 1 & 0.91 & 0.05 & 0.5  & 0.05 & 0.05\\
		8  & 1 & 0.88 & 0.09 & 0.49 & 0.1 & 0.09\\
		\hline
	\end{tabular}
\caption{Inductively coupled plasma mass spectrometry analysis of chemical composition of Sr$($Co$_{1-x}$Ni$_{x})_2$As$_2$ single crystals used in our neutron
scattering experiments. We have carried out measurements on 8 samples. After normalizing the measured concentrations of Co, Ni, and Sr to the 100\% As concentration, we find that 
the nominal Ni-doping is fairly close to the actual Ni-doping.}
\end{center}
\end{table}
\endgroup

Our time-of-flight (TOF) inelastic neutron scattering (INS) experiments were carried out at the fine-resolution Fermi chopper spectrometer (SEQUOIA) and wide Angular-Range Chopper Spectrometer (ARCS) at the Spallation Neutron Source (SNS), Oak Ridge National Laboratory (ORNL). The neutron diffraction experiments were done on the HB-3A four-circle diffractometer at the High Flux Isotope Reactor (HFIR), Oak Ridge National Laboratory \cite{Cao} and high resolution powder diffractometer - BT-1 at the NIST Center for Neutron Research. To facilitate easy comparison with INS results of SrCo$_2$As$_2$ \cite{SCA_ly}, We define the momentum transfer {\bf Q} in three-dimensional reciprocal space in ${\rm \AA^{-1}}$ as ${\bf Q} = H{\bf a}^* + K{\bf b}^* + L{\bf c}^*$, where $H$, $K$, and $L$ are Miller indices and ${\bf a}^* = {\bf \hat{a}}2\pi/a$, ${\bf b}^* = {\bf \hat{b}}2\pi/b$, and ${\bf c}^* = {\bf \hat{c}}2\pi/c$ with $a = b \approx 5.614$ ${\rm \AA}$, and $c = 11.566$ \AA. For TOF experiments, our single crystals were co-aligned in the $[H, 0, L]$ scattering plane. At $T = 5$ K, the incident beam with energies 
of $E_i = 12, 30,$ and 80 meV is parallel to the $c$-axis of the crystals. In neutron diffraction experiments on HB-3A, neutron wavelength of 1.553 \AA\ was
used from a bent perfect Si (2,2,0) monochromator. A closed-cycle refrigerator was used to provide the temperature above 4 K. The aluminum pins were used 
to mount the crystals.

To study the changes of the helical magnetic structure with $x$ in Sr$($Co$_{1-x}$Ni$_{x})_2$As$_2$, we carried out neutron diffraction measurements to investigate the $x$-dependence of the incommensurate magnetic ordering wave vector $(0,0,q)$ along the $c$-axis [Figs. 1(c) and 1(d)]. Figures 2(a-d) show 
detailed scans along the $[0, 0, L]$ direction 
at $T = 5$ K for $x = 0.05, 0.125, 0.15$, and 0.2. These scans reveal incommensurate peaks at (0, 0, $n \pm q$), with $n$ being an even integer and $q$ = 0.6, 0.51, 0.48, and 0.38 for $x$ = 0.05, 0.125, 0.15 and 0.2, respectively. Figures 2(e-h) show the temperature dependence of the magnetic order parameter for this series of Sr$($Co$_{1-x}$Ni$_{x})_2$As$_2$ compounds, where the results of $x = 0.1$ is reported in Ref. \cite{SCA_ly}.

Figure 1(e) shows the Ni-doping evolution of $(0,0,q)$ and the magnetic transition temperature $T_N$ in Sr(Co$_{1-x}$Ni$_x$)$_2$As$_2$. While $T_N$ has a maximum 
at $x = 0.1$, the magnetic propagation vector $q$ monotonously decreases from $q = 0.6$ in $x=0.05$ to $q = 0.38$ in $x=0.2$. According to the 1D frustrated Heisenberg Hamiltonian with the easy-plane anisotropy \cite{Helical_theory,Helical_theory1,Helical_theory2,Helical_theory3}, $q < 0.5$ means that the turn angle between the moments in neighboring layers are smaller than 90$^\circ$ and thus the effective inter-layer exchange coupling is FM \cite{Helical,SCNA_ames,SCNA_ly}. Although it is unclear that $q$ will reduce to zero for the sample with $T_N = 0$, its magnetic properties are likely dominated by FM spin fluctuations. In the 1D Heisenberg Hamiltonian, the presence of a next-nearest-neighboring (NNN) AF coupling $J_{c2}$ 
and its competition with the FM inter-layer coupling $J_{c1}$ induce a helical magnetic order. In the limit of $J_{c2} \rightarrow 0$, Sr(Co$_{1-x}$Ni$_x$)$_2$As$_2$ will
become FM. The small but finite AF $J_{c2}$ causes ripples in the uniform FM background
of $J_{c1}$, thus inducing the helical magnetic order in 
Sr(Co$_{1-x}$Ni$_x$)$_2$As$_2$ by an order-by-disorder mechanism through 
the RKKY interactions \cite{QOBD}.

\begin{figure}[t]
\includegraphics[scale=.45]{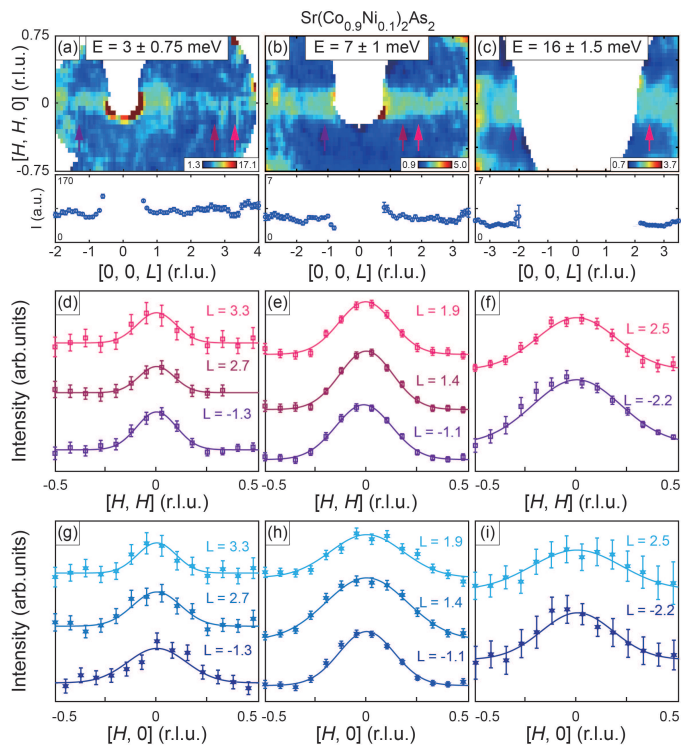}
\caption{(a,b,c) Constant energy slices of spin fluctuations of Sr$($Co$_{0.9}$Ni$_{0.1})_2$As$_2$ within the $[$H, H, L$]$ plane at $E = 3 \pm 0.75, 7 \pm 1$, and $16 \pm 1.5$ meV, respectively. The corresponding 1D cuts along the $[0, 0, L]$ direction are shown in the bottom panel. (d-f) 1D cuts of spin fluctuations 
along the $[H, H]$ direction at different $L$ and energies. (g-i) Corresponding 1D cuts along the $[H,0]$ direction. The solid lines are Gaussian fits to the cuts.
}
\end{figure}

To explore spin fluctuations in Sr(Co$_{1-x}$Ni$_x$)$_2$As$_2$, we performed time-of-flight (TOF) neutron scattering experiments on single crystals of $x=0$ \cite{SCA_ly} and $x=0.1$. Figures 3(a-d) and (i-l) show images of spin fluctuations in the $[H,K]$ plane for $x=0.1$ and 0, respectively, at $E = 12\pm 2$, $17\pm 3$, $35\pm 5$, 
and $52\pm 7$ meV. The stripe-type AF fluctuations are clearly seen at $(0,1)/(1,0)$ positions in both compounds and are longitudinally elongated, forming ridge-like excitations at higher energies. On the other hand, FM spin fluctuations near $(0,0)$ are greatly enhanced in the $x=0.1$ compared to those in $x=0$, as confirmed by
neutron polarization analysis \cite{SCA_ly}. 
For comparison, the spin dynamic susceptibilities calculated by the DFT+DMFT method
at energies similar to experiments are plotted in Fig. 3(e-h).
The DFT+DMFT calculations are identical as those 
described in previous work \cite{SCA_ly}.  
The only difference is that the Ni-doping is approximated by a virtual crystal approximation. For example, for 10\% Ni doping, a virtual Co atom was used, with a nuclear charge 27.1 and a number of 27.1 electrons. 
While the overall results of the calculations are consistent with the experimental data, the calculations indicate much stronger FM intensity \cite{ACS_zhiping}.

To understand the magnetic exchange modulations along the $c$-axis in Sr(Co$_{1-x}$Ni$_x$)$_2$As$_2$, the $L$ dependence of the FM spin fluctuations has to be measured.
In Figures 4(a-c), we plot 2D images of 
the background subtracted scattering intensities of the $x=0.1$ compound 
in the $[H,H,L]$ plane at $T = 5$ K for energies of $E = 3\pm 0.75$, $7\pm 1$, 
and $16\pm 1.5$ meV. The 2D nature of FM magnetic fluctuations is clearly seen as rod-like scattering without $L$ modulation, as confirmed by   
the corresponding 1D cuts along the $[0,0,L]$ direction [Figs. 4(d-i)].
Since the $x=0.1$ compound has a $c$-axis helical magnetic ordering wave vector   
at $q = 0.56$, we expect critical scattering and low-energy spin
waves to stem from this wave vector \cite{SCNA_ly}. However, we find no evidence of the expected 3D magnetic excitations down to 2 meV around $q = 0.56$, much different 
from the spin excitations associated with the 3D AF order in NaFeAs \cite{NFA}. 
On the other hand, the widths of the rod-like magnetic scattering are broadened 
with increasing energy [Figs. 4(a-c)]. These are attributed to the sharp dispersion of magnetic excitations along the $[H,H]$ direction, indicating extremely anisotropic exchange couplings. In Figs. 4(d-i), we show a series of 1D cuts of magnetic excitations along the $[H,H]$ and $[H,0]$ directions at different $L$ values and energies. The FM correlations within the CoAs plane are isotropic at all measured energies.

\begin{figure}[t]
\includegraphics[scale=.48]{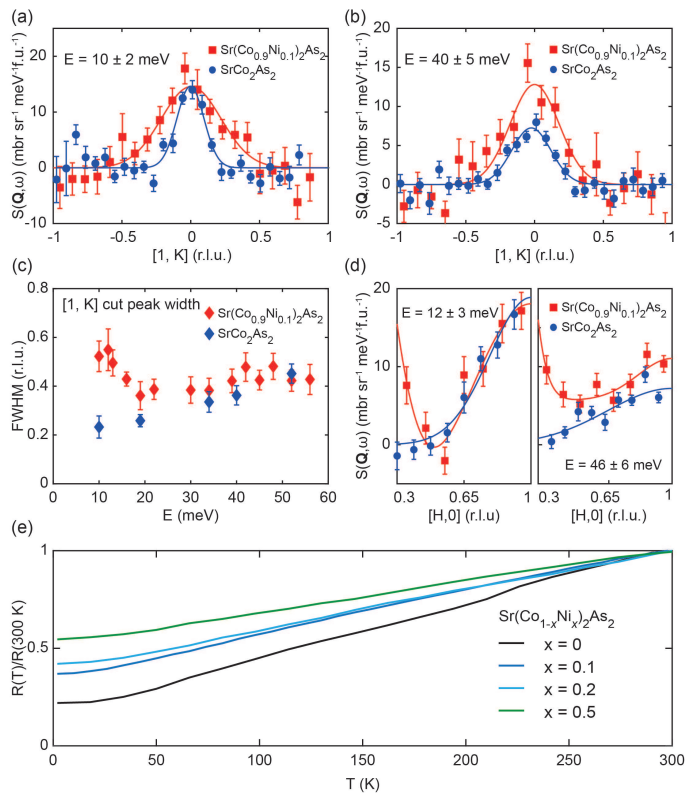}
\caption{(a,b) Transverse cuts of the AF spin fluctuations in Sr(Co$_{1-x}$Ni$_x$)$_2$As$_2$ with $x=0,0.1$ at $E = 10\pm 2$ and $40\pm 5$ meV, respectively.
The solid lines are Gaussian fits to the data. 
(c) Energy dependence of the line widths of AF spin 
fluctuations along the $[1,K]$ direction.
(d) The longitudinal scans of the spin fluctuations in the $x=0$ and 
0.1 at $E = 12\pm 3$ and $46\pm 5$ meV. (f) Temperature dependence
of the normalized electrical resistivity $\rho_c(T)/\rho_c(300\ {\rm K})$ along
the $c$-axis for $x=0, 0.1, 0.2,$ and 0.5. 
}
\end{figure}

To further understand the evolution of spin fluctuations in $x=0$ and 0.1, we compared a series of transverse cuts of
the stripe-type AF spin fluctuations in these compounds. 
Figures 5(a) and 5(b) show Gaussian fits for cuts
along the $[1,K]$ direction 
at $E = 10\pm 2$ and $40\pm 5$ meV, respectively. 
While spin excitations at 10 meV are similar for both compounds, 
the peak in $x=0.1$ is much broader. In CaCo$_{2-y}$As$_2$, there are strong frustrations between the NN exchange coupling $J_1$ and NNN exchange coupling $J_2$ within the Co square lattice. The frustration leads to the dimensionality reduction and hence the 2D Co lattice is described by an array of nearly decoupled 
FM spin chains \cite{CCA1}. From this perspective, the peak widths of the $[1,K]$ scans 
at the $(0,1)/(1,0)$ positions represent the inverse correlation length along the 1D spin chain. The broad line width of $x=0.1$ sample at 10 meV in Fig. 5(a) indicates a short magnetic correlation length along the FM spin chain, and is likely due to the disorder introduced by Ni substitution. As the energy increases, such a broadening effect is reduced and observed to  $\sim$20 meV [Fig. 5(c)]. The shorter spin correlation length caused by disorder disfavors the stripe AF alignment of the magnetic moments and thus supports the FM order within the plane. This is consistent with the observation that the 2D FM layers are formed in Sr(Co$_{1-x}$Ni$_{x})_2$As$_2$ in spite of the fact that FM fluctuations are considerably weaker than those in CaCo$_2$As$_2$ \cite{ACS_zhiping,CCA1}. At 40 meV, the line widths of the $x=0$ and 0.1 compounds are basically the same.  There is a slight peak broadening with increasing energy in both compounds, which is due to the sharp dispersion of spin excitations along the 1D spin chain [Fig. 5(c)]. Figure 5(d) shows the longitudinal cuts of magnetic excitations along the $[H,0]$ direction at $E = 12$ and 46 meV, respectively, demonstrating the large FM contribution in the spin excitation spectra induced by Ni substitution.

\begin{figure}[t]
    \centering
    \includegraphics[scale=0.61]{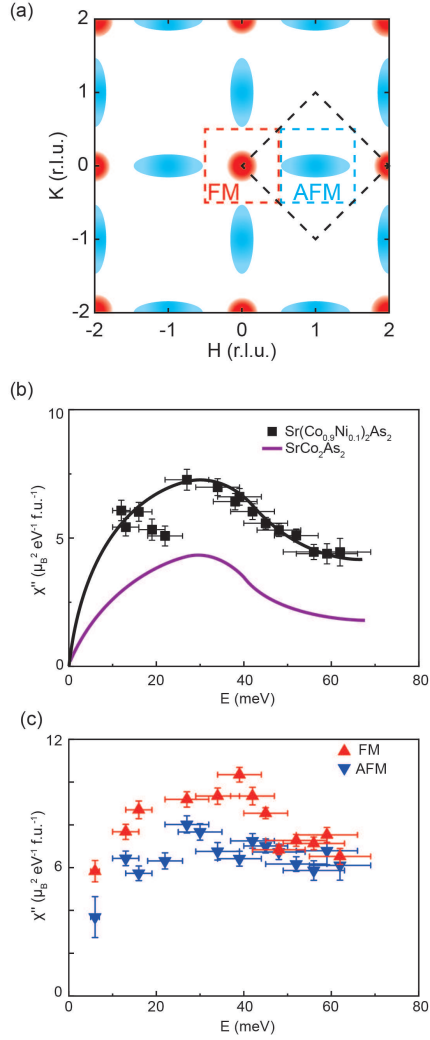}
    \caption{(a) Schematics of spin fluctuations in the $[H, K]$ two-dimensional reciprocal space, where the red (blue) areas represent FM (AF) spin fluctuations. The black diamond marks the integration region for calculating energy-dependent local dynamic susceptibility. The red and blue squares are the the integration region for calculating energy-dependent FM and AF local dynamic susceptibility. (b) Energy-dependent local susceptibility $\chi^{\prime\prime}(E)$ of Sr$($Co$_{0.9}$Ni$_{0.1})_2$As$_2$ in absolute units normalized by using a vanadium standard. The black line is a guide to the eye. The purple line is the result of SrCo$_2$As$_2$ from our previous work \cite{SCA_ly}. (c) Energy dependence of the integrated $\chi^{\prime\prime}(E)$ of FM (red) and AF (blue) spin fluctuations of Sr$($Co$_{0.9}$Ni$_{0.1})_2$As$_2$.}
    \label{fig:my_label}
\end{figure}

Using elastic and inelastic neutron scattering, we have demonstrated that Sr(Co$_{1-x}$Ni$_{x})_2$As$_2$ is close to a FM instability and hosts strong 2D FM spin fluctuations. However, Ni-doped Sr(Co$_{1-x}$Ni$_{x})_2$As$_2$ has a helical order of the stacked FM layers along the $c$-axis, meaning that the inter-layer magnetic exchange coupling cannot be purely FM \cite{SCNA_ly,SCNA_ames,Wilde2019}. From the in-plane electrical  resistivity measurements on Sr(Co$_{1-x}$Ni$_{x})_2$As$_2$, it is known that the resistivity is essentially Ni-doping 
independent at room temperature but increases with increasing Ni-doping at low-temperature \cite{SCNA_ames}.       
In contrast to FM La-doped SrCo$_2$As$_2$ in which the La is located out of the CoAs plane, Ni-substitution directly causes disorder in the magnetic Co lattice and
therefore induces an in-plane resistivity increase with increased doping. 
However, since Ni-doping causes a decreased $c$-axis lattice constant and inter-layer
distance \cite{SCNA_ames}, it is unclear if doping will increase or decrease 
electric resistivity along the $c$-axis. Figure 5(e) shows normalized $c$-axis electrical resistivity $\rho_c(T)/\rho_c(300\ {\rm K})$ for
Sr(Co$_{1-x}$Ni$_{x})_2$As$_2$ crystals with $x = 0, 0.1, 0.2, 0.5$. 
 $\rho_c(T)/\rho_c(300\ {\rm K})$ increases with increasing Ni-doping at low temperatures, suggesting that such effect is also determined mostly by substitutional disorder in spite of the reduced layer distances.

To further study the evolution of spin fluctuations in Ni-doped SrCo$_2$As$_2$, we consider the energy dependence of the local spin dynamic susceptibility, 
defined as $\chi^{\prime\prime}(E) = \int_{\rm BZ} \chi^{\prime\prime}({\bf Q},E)d{\bf Q}/\int_{\rm BZ}d{\bf Q}$, where $\chi^{\prime\prime}({\bf Q},E)$ is the wave vector and energy dependence of the imaginary part of the dynamic susceptibility within a Brillouin zone (BZ) [black diamond in Fig. 6(a)] \cite{Dai2015}. Figure 6(b) shows the comparison of the energy dependence of the local dynamic susceptibility of Sr$($Co$_{0.9}$Ni$_{0.1})_2$As$_2$ and SrCo$_2$As$_2$. Figure 6(c) shows the energy dependence of integrated FM (red) and AF (blue) spin fluctuation intensities for Sr$($Co$_{0.9}$Ni$_{0.1})_2$As$_2$, where the corresponding integration region is marked in Fig. 6(a).

\section{Discussion}

To understand these observations, we first consider if the helical order can be induced by Fermi surface nesting. Figures 1(f) and 1(g) show the 
electronic band structure and Fermi surfaces at $k_z=0$, respectively, obtained by DFT+DMFT calculations for $x=0.1$. The 3D Fermi surfaces for different orbitals are shown in Figs. 1(h,i). Considering the large variation of $q$ with a relatively small amount of electron doping [Fig. 1(e)], the band dispersion related to $q$ has to be quite flat so that a small change of chemical potential $\mu$ will lead to a large variation in the Fermi surface. This may be related to the flat portion of the $d_{x^2-y^2}$ band around the $M$ point [Fig. 1(f)], where its hybridization 
with the $d_{z^2}$ orbital adds a finite $k_z$ dispersion. 
Due to the small energy scale, this effect is not captured accurately by the calculation. However, the variation of the 3D Fermi surface along the 
diagonal direction ($\Gamma-M$) from $x=0$ \cite{ACS_zhiping} to $x=0.1$ 
[Fig. 1(i)] qualitatively agrees with the observation that the magnetic propagation vector decreases with increasing Ni-doping. 
Nevertheless, there is no evidence of Fermi surface
nesting induced helical spin structure [Figs. 1(c-e)].

Alternatively, if we assume that the NNN inter-layer coupling $J_{c2}$ cannot arise from the direct exchange coupling due to the large layer separation, the itinerant 
electron-induced RKKY-type oscillating magnetic interaction may be responsible
for the dominant $J_{c2}$ rather than $J_{c1}$ in the intermediate doped compounds with $q \sim 0.5$. In the simplest case of the RKKY model, the magnetic exchange coupling is oscillating as a function of $J({\bf Q})$ in which ${\bf Q} = 2{\bf k}_F$ (where ${\bf k}_F$ is the nesting wave vector). This mechanism will give a magnetic ordering wave vector from RKKY interaction similar to a nesting picture. However, the situation is different when there are multiple Fermi surfaces with different nesting wave vectors. The flat band at the Fermi level may also contribute to the RKKY interaction. Consider a simple case in which there are two competing nesting wave vectors ${\bf Q}_1$ and ${\bf Q}_2$.  The frustration between ${\bf Q}_1$ and ${\bf Q}_2$ will lead to a magnetic order at a third wave vector ${\bf Q}_3$ that is different from ${\bf Q}_1$ and ${\bf Q}_2$, and thus cannot be explained by a nesting picture. Therefore, the absence of a nesting vector
 does not imply the failure of the RKKY mechanism, but actually suggests that the frustration (either in the real space or reciprocal space) plays a significant role in the formation of a helical magnetic order. 
In this picture, the Ni-doping induced disorder only 
affects low energy spin fluctuations, and causes a small change in the effective couplings $J_{c1}$ and $J_{c2}$ through the RKKY interactions that stabilize the helical modulation of the magnetic moments near a FM instability \cite{FMQCP1,FMQCP2}.
While the RKKY interaction induced magnetic frustration may be able to explain the
observed helical order, we cannot rule out the possibility that the helical order arises from the Fermi surface nesting not captured by the DMFT calculation.

Recently, intriguing correlation effects were found to arise in partially flat-band systems comprised of the flat and dispersive portions in the same region of the reciprocal space \cite{fb1,AFB,SRO_FM,SRO_SC}. When electrons are filling the flat portion of the band, non-Fermi liquid behavior arises even for systems with intermediate electron-electron interactions. Sr(Co$_{1-x}$Ni$_{x})$As$_2$, which has a saddle-like flat band at the Fermi level and intermediate electron 
correlations \cite{SCNA_ly,SCNA_ames,Wilde2019}, provides a platform to study the correlation physics in partially flat-band system and offers versatile controls via 
electron/hole doping and band engineering.

In conclusion, we used neutron scattering to study the doping dependence of the magnetic wave vector of the helical order and spin fluctuations of Sr(Co$_{1-x}$Ni$_{x})_2$As$_2$. We find that the magnetic wave vector monotonously decrease with Ni-doping. The spin fluctuations in the probed range are 2D like at $x=0.1$ in spite of the underlying helical magnetic order. While we find no evidence of Fermi surface nesting induced helical magnetic order based in DFT+DMFT calculation, our results are consistent with an order-by-disorder phase transition induced by the itinerant electron mediated RKKY interactions.

\section{Acknowledgements}

Single crystal growth and the work for neutron scattering were supported by US NSF Grant No. DMR-1700081 and Robert A.
Welch Foundation Grant No. C-1839 (P.D.). Z.P.Y. was supported by the NSFC (Grant No. 11674030), the Fundamental Research Funds for the Central Universities (Grant No.
310421113), and the National Key Research and Development Program of China, Grant No. 2016YFA0302300. The calculations used high performance computing clusters at
BNU in Zhuhai and the National Supercomputer Center in
Guangzhou. Neutron diffraction experiment at NIST were supported by the US Department of Energy under EPSCoR Grant No. DE-SC0012432 with additional support from the Louisiana Board of Regents. A portion
of this research used resources at the Spallation Neutron
Source and High flux Isotope Reactor, DOE Office of Science User Facilities operated
by the Oak Ridge National Laboratory.

\end{document}